\documentclass[reprint,superscriptaddress,longbibliography, amsmath,amssymb,aps,prb]{revtex4-2}

\usepackage{amsmath}
\usepackage{physics}
\usepackage[colorlinks=true,citecolor=blue]{hyperref}
\usepackage{graphicx}% Include figure files
\graphicspath{{Pictures/}}
\usepackage{bm}% bold greek math
\usepackage[dvipsnames]{xcolor}
\usepackage[normalem]{ulem}
\usepackage{gensymb}% Special symbols

\begin{document}

\preprint{APS/123-QED}

\title{Ultrafast inertia-free switching of double magnetic tunnel junctions}

\author{Yu. Dzhezherya}
\affiliation{Institute of Magnetism of the NAS of Ukraine and MES of Ukraine, 03142 Kyiv, Ukraine}%

\author{P. Polynchuk}
\affiliation{Institute of Magnetism of the NAS of Ukraine and MES of Ukraine, 03142 Kyiv, Ukraine}%

\author{A. Kravets}
\affiliation{Nanostructure Physics, Royal Institute of Technology, 10691 Stockholm, Sweden}%
\affiliation{Institute of Magnetism of the NAS of Ukraine and MES of Ukraine, 03142 Kyiv, Ukraine}%

\author{V. Korenivski}
\affiliation{Nanostructure Physics, Royal Institute of Technology, 10691 Stockholm, Sweden}%

\begin{abstract}
We investigate the switching of a magnetic nanoparticle comprising the middle free layer of a memory cell based on a double magnetic tunnel junction under the combined effect of spin-polarized current and weak on-chip magnetic field. We obtain the timing and amplitude parameters for the current and field pulses needed to achieve 100 ps range \emph{inertia-free} switching under \emph{least-power} dissipation. The considered method does not rely on the stochastics of thermal agitation of the magnetic nanoparticle typically accompanying spin-torque switching. The regime of ultimate switching speed-efficiency found in this work is promising for applications in high-performance nonvolatile memory.  
\end{abstract}

\maketitle

Slonczewski and Berger \cite{Slon96,Berger96,Slon02,Berger02} proposed a mechanism of excitation and switching in magnetic nanostructures based on spin-transfer-torques (STT), which was demonstrated experimentally \cite{Katine00,Kiselev03} and found wide use in magnetic random access memory (MRAM). STT-based MRAM has become a successful technology, offering fast, non-volatile, radiation and thermally hard (can operate at up to 400\degree C) data storage in various industrial applications such as mobile, automotive, military, space \cite{Thomas14,Kent15}. 

More recently, MRAM based on double magnetic tunnel junctions (DMTJ) was developed to enhance STT by spin injection from both top and bottom magnetic junctions \cite{Worledge22,Worledge22a}. The two fixed reference layers in a DMTJ are oriented antiparallel, such that the combined STT is a sum of the individual-MTJ spin-torques. The two junctions are made to have different resistance for readout efficiency (higher magneto-resistance, MR). An alternative design, most promising today, the Double Spin-torque MTJ (DSMTJ) has one of the oxide junctions replaced with a nonmagnetic metallic spacer, which does not contribute to the total resistance and hence does not dilute the MR. The analysis presented in this paper is directly relevant to both DMTJ and DSMTJ MRAM.

Worledge \cite{Worledge17} recently developed a single-domain model for switching of a double magnetic tunnel junction (DMTJ) magnetized perpendicular to the plane, which is a very promising implementation of MRAM since it significantly increases the strength of available STT. The obtained analytical result for the threshold switching current showed a reduction by up to 10-fold compared with single-MTJ based cells.

In this work, we theoretically investigate the important technological aspect of how to achieve ultimate memory writing speed while expending least amount of energy; more specifically, how to most efficiently combine STT and word/bit-line Oe-fields available on-chip. The role of a weak Oe-field added to strong STT is to achieve a minor magnetization non-collinearity in the DMTJ stack, which significantly improves the STT efficiency during switching\footnote{We note that a minor magnetization non-collinearity between the fixed and free layers of the DTMJ (DSMTJ) can in practice be achieved by setting the exchange-pinning direction in the fixed layers slightly off the $x$ axis. Such implementation would not require the assisting Oe-field pulse and the writing of the memory cell would be done solely by STT.}. We will show that an optimal combination of STT and Oe-field pulses can be determined analytically for given device parameters and leads to inertia-free switching in the 200 ps range. 

\begin{figure}[h]
\includegraphics[width=85mm]{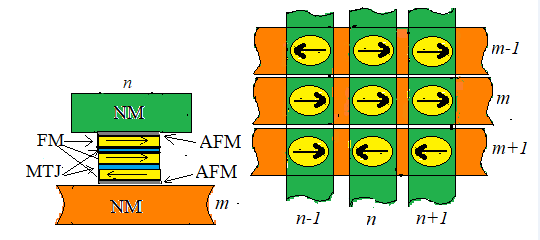}
\caption{Schematic representation of MRAM memory cell (left) consisting of three magnetic nanoparticles (arrows indicate directions of magnetization) located between crossed bit (n) and word (m) lines. Planar layout of memory array (right) showing positions of FM free layers centered at bit-word line cross points.}
\label{fig:mesh1}
\end{figure}

We consider a DMTJ-based MRAM cell, where the current through the in-plane magnetized junction, spin-polarized by the magnetically co-linear and antiparallel fixed magnetic bottom and top layers, exerts a spin-torque on the middle free layer, which can be arbitrary angled in the plane by a field from the word or bit line of the memory array, as illustrated in Fig. \ref{fig:mesh1}.

We use the Landau-Lifshitz-Slonczewski equation \cite{Slon96,Slon02}, which takes into account the spin transfer torque (STT) of the spin-polarized current through the stack. In order to simplify the theoretical consideration while focusing on the main idea of the paper, we will neglect the dissipation in the system since it leads to an insignificant quantitative modification of the final results for the considered process. We then have
\begin{equation}
\frac{d \vb*{m}} {dt} = -\gamma[\vb*{m} \cp \vb*{H}_{\mathrm{eff}}] -\gamma 4\pi M_{s}\kappa[\boldmath{m}\times[\boldmath{m}\times\boldmath{\mu}]],
\end{equation}
where $\vb*{m}$ is the unit magnetization vector of the free layer, which is the functional element of the system, $\vb*{H}_{\mathrm{eff}}$ is the effective magnetic field acting on the magnetic moment of the free layer, \(\gamma=\frac{2\mu_{\mathrm{B}}}{\hbar}\) is the gyromagnetic ratio, \(\kappa=\frac{\hbar\cdot V\cdot P}{{R}_{\mathrm{\perp}}\cdot2e\cdot4\pi M_{\mathrm{s}}^2\cdot S\cdot d}\) is a normalised coefficient that determines the intensity of STT, $V$ is the potential difference between the bit and word lines, \(R_{\perp}\) is the total electrical resistance of DMTJ, $P$ is the polarization coefficient of the current through the free layer, $\vb*{M}_{\mathrm{s}}$ is the saturation magnetization of the free layer, and $\vb*{\mu}$ is the unit magnetization vector of the bottom fixed layer. In turn, $S$ and $d$ are the area and thickness of the free layer, respectively.

As in \cite{Che11,Dzhe13,Koop17}, we assume that the free layer has the shape of a thin ellipse with a small in-plane eccentricity, which determines its shape anisotropy. The strong out-of-plane demagnetization and weak in-plane anisotropy confines the magnetization to be strongly in-plane and along the long semi-axis of the ellipse at equilibrium.

Formally, this system is similar to the one considered in \cite{Worledge17}, with the difference being perpendicular orientation of the magnetic moments in the previous work, unlike the in-plane orientation in our case. However, the angular dependence of the STT coefficient \(\kappa\) in both systems is similar. In the case of small spin-polarization coefficient, $P \ll 1$, the angular dependence of $\kappa$ is manifest only in the terms proportional to $P^3$. Hence, with good accuracy, $\kappa$ can be taken as angle-independent\footnote{In practice, the first nonlinear term ($P^3$) is small for all $P \lesssim 0.5$, which is the common MTJ polarization range in STT-MRAM\cite{Worledge17}}.

It should be noted that a single fixed magnetic layer (bottom or top) is in principle sufficient to control the magnetization direction of the free layer. STT due to spin-polarized electrons transmitted from the free to fixed layer favors the antiparallel state of the MTJ, while the reversed current favors the parallel state. Thus, bi-directional switching is possible, although sub-optimal vis-avi DMTJ, as we detail below.  

The three-layer DMTJ structure illustrated in Fig. \ref{fig:mesh1} has a significant operational advantage since the magnetization switching process is symmetric due to the mutual compensation of the dipolar stray fields form the bottom and top fixed magnetic layers. In addition, as in the system considered in \cite{Worledge17}, the two STT contributions add up and enhance the effective torque acting on the magnetic moment of the free layer. In what follows, we therefore use coefficient $P$ as taking into account the STT contributions from both junctions of the DMTJ and neglect the stray fields from the outer fixed layers.

For this system to operate as a memory cell with a resistive readout, it is sufficient that the (magneto-) resistances of the two tunnel junctions are different.

The flat shape of the free layer results in strong easy-plane magnetic anisotropy, so we can take $|m_z| \ll 1$. Accurate to linear terms in $m_z$, the unit magnetisation vector of the free layer is then
\begin{equation}
    \vb*{m}=\frac{\vb*{M}}{\vb*{M_{\textrm{s}}}}=(\cos(\phi), \sin(\phi), m_{z});
\end{equation}
with that of the bottom fixed layer  taken as
\begin{equation}
\vb*{\mu}=(-1, 0, 0).
\end{equation}
The effective field, $\vb*{H}^{i}_{\textrm{eff}}$, which takes into account demagnetization and the external magnetic field, is given by
\begin{equation*}
    {H}^{x}_{\textrm{eff}}=-4\pi M_{\textrm{s}}N_{x}cos(\phi)+H_{x}
\end{equation*}
\begin{equation}
  {H}^{y}_{\textrm{eff}}=-4\pi M_{\textrm{s}}N_{y}sin(\phi)+H_{y}  
\end{equation}    
\begin{equation*}    
    {H}^{z}_{\textrm{eff}}=-4\pi M_{\textrm{s}}N_{z}m_{z}
\end{equation*}

After substituting (2-4), we obtain the Landau-Lifshitz equation in new variables, which, after certain simplifications, take the form:
\begin{equation*}
    \frac{d\phi}{d\tau}=-m_{z}+\kappa \sin(\phi)
\end{equation*}
\begin{equation}    
    \frac{dm_{z}}{d\tau}=(N_{y}-N_{x}) \sin(\phi) \cos(\phi) -h_{y} \cos(\phi)+h_{x} \sin(\phi),
\end{equation}
where $\tau=t\cdot\omega_{\textrm{0}}$, $\omega_{\textrm{0}}=8 \pi M_{\textrm{s}} \mu_{\textrm{B}}/\hbar$, $\mu_{B}$ is Bohr's magneton, 
$h_{y}=\frac{H_y}{4\pi M_{\textrm{s}}}$, $h_{x}=\frac{H_x}{4\pi M_{\textrm{s}}}$.

In (5) it was taken into account that \(\lvert m_{z}\rvert\), \(N_{y}\), \(N_{x}\), \(\lvert h_{y}\rvert\), \(\lvert h_{x}\rvert\), $\kappa \ll 1$, \(N_{z}=1-N_{y}-N_{x}\approx1\), and terms no higher than linear in the small parameter were retained. 

This pair of equations describing the behaviour of the magnetization of a free layer is equivalent to a single second-order equation for the angular variable:
\begin{equation}
\begin{aligned}
\ddot{\phi}+(N_{y}-N_{x}) \sin(\phi) \cos(\phi)-\dot{\kappa} \sin(\phi)-\kappa \cos(\phi)\dot{\phi} \\ -h_{y} \cos(\phi)+h_{x} \sin(\phi)=0.
\end{aligned}
\end{equation}

Hereinafter, a point above the physical quantities will be used to denote the time derivative: \(\dot{\phi}=\frac{d\phi}{d\tau}\), \(\dot{\kappa}=\frac{d\kappa}{d\tau}\)... Equation (6) is used below as the basis for developing a theory of controlling the magnetic states of a DMTJ memory cell, with a focus on ultra-high speed and relaxation-free switching.

Let us first consider the case when there is no external field \(h_{y}=h_{x}=0\), but the magnetic moment of the free layer is subjected to a spin-polarized current. We assume that the current pulse has a rectangular time dependence \(\kappa(\tau)=\kappa_{0}\Theta(\tau+\frac{T\omega_{0}}{2})\cdot\Theta(\frac{T\omega_{0}}{2}-\tau)\), where \(\Theta(\tau)\) is the Heaviside function and \(\frac{T\omega_{0}}{2}\) is a dimensionless coefficient that includes the pulse duration $T$. The rectangular shape of the pulse means that the rise time of the current front is much shorter than the characteristic response time of the system to the field and current excitation. We will determine the exact value of the characteristic time later.

It is obvious that such time dependence results in the spin polarization coefficient changing as \(\dot{\kappa}=\kappa_{0}(\delta(\tau+\frac{T\omega_{0}}{2})-\delta(\tau-\frac{T\omega_{0}}{2}))\), where \(\delta\) is the Dirac delta function. Thus, the solution of equation (6) can be decomposed into three sequential dependencies: \(\phi_{I}\) in the interval $\left(-\infty,-\frac{T\omega_{0}}{2} \right)$, \(\phi_{II}\) in $\left[-\frac{T\omega_{0}}{2},\frac{T\omega_{0}}{2} \right]$ and \(\phi_{\textrm{III}}\) in $\left(\frac{T\omega_{0}}{2},+\infty\right)$.

The boundary conditions are determined by integrating equation (6) in the vicinity of points \(\tau_{1}=-\frac{T\omega_{0}}{2}\) and \(\tau_{2}=\frac{T\omega_{0}}{2}\). Therefore, taking into account \(\boldmath{h=0}\), we have:
\begin{equation*}
\ddot{\phi}_{I}+(N_{y}-N_{x})sin(\phi_{I})cos(\phi_{I})=0,
\end{equation*} 
with  $\tau\in \left(-\infty,-\frac{T\omega_{0}}{2}\right)$;
\begin{equation}
\ddot{\phi}_{\textrm{II}}+(N_{y}-N_{x})sin(\phi_{\textrm{II}})cos(\phi_{\textrm{II}})-\kappa_{0}cos(\phi_{II})\dot{\phi}_{II}=0,
\end{equation}
with $\tau \in \left[-\frac{T\omega_{\textrm{0}}}{2},\frac{T\omega_{\textrm{0}}}{2}\right]$;
\begin{equation*}
\ddot{\phi}_{\textrm{III}}+(N_{y}-N_{x})sin(\phi_{\textrm{III}})cos(\phi_{\textrm{III}})=0,
\end{equation*}
with $\tau \in \left(\frac{T\omega_{\textrm{0}}}{2},+\infty\right)$. The boundary conditions are
\begin{equation*}
\phi_{\textrm{I}}=\phi_{\textrm{II}};  \dot{\phi}_{\textrm{II}}-\dot{\phi}_{\textrm{I}}=\kappa_{0}sin(\phi_{\textrm{I}}),  \text{with } \tau=\tau_{1}=-\frac{T\omega_{0}}{2}
\end{equation*}
\begin{equation}
\phi_{\textrm{II}}=\phi_{\textrm{III}}; \dot{\phi}_{\textrm{III}}-\dot{\phi}_{\textrm{II}}=-\kappa_{0}sin(\phi_{\textrm{II}}),  \text{with } \tau=\tau_{2}=\frac{T\omega_{0}}{2},
\end{equation}
If we assume that before the current was switched on, the magnetic moment of the free layer was in state of equilibrium, then the following boundary condition will be fulfilled at \(\tau=\tau_{1}=-\frac{T\omega_{0}}{2}\): \(\phi_{\textrm{I}}=\phi_{\textrm{II}}=0\), \(\dot{\phi}_{\textrm{II}}=0\). Thus, the problem describing the perturbation of the magnetization of the free layer under the excitation of the spin-polarized current pulse is as follows:
\begin{equation}
\ddot{\phi}_{\textrm{II}}+(N_{y}-N_{x}) \sin(\phi_{\textrm{II}}) \cos(\phi_{\textrm{II}})-\kappa_{0} \cos(\phi_{\textrm{II}})\dot{\phi}_{\textrm{II}}=0,
\end{equation}
\begin{equation*}
\phi_{\textrm{II}}=0; \dot{\phi}_{\textrm{I}}=0 \text{ for } \tau=-\frac{T\omega_{0}}{2}.
\end{equation*}

Solving this system shows that \(\phi_{\textrm{II}}(\tau)=0\). Similarly, \(\phi_{\textrm{III}}(\tau)\equiv0\). This means that in a collinear system, when the magnetic moments of the pinned and free layers are parallel, a spin-polarised current does not affect the magnetization unless there are additional factors that violate the collinearity condition. Thermal fluctuations of the magnetisation vector can be one of these factors, but in this case, the control process is randomized and results in rather significant parameter spreads for writing and reading out information.

In order to avoid random processes, we consider a memory cell design where a weak magnetic field pulse \(h_{y}\) is added to the write sequence. This breaks the collinearity of the magnetisation of the free and pinned layers. The source of such a field can typically be the flat current bus passing above/below a given memory cell in the Ox/Oy direction (word/bit line of the memory array). Based on Maxwell's equation $\curl \bm{H}=4\pi \bm{j}/c$), it is easy to show that the current flowing in, say, the word-line induces a magnetic field $H_{y}$, which at a distance much smaller than the line width is equal to  (in SGS units) $H_{y}=\frac{2\pi I_{x}}{cL}$, or $h_{y}=\frac{I_{x}}{2cM_{\textrm{s}}L}$, where $c$ is the speed of light, $L$ is the word-line width, and $I_{x}$ is the current flowing through the line in the Ox direction.

\begin{figure}[h]
\includegraphics[width=85mm]{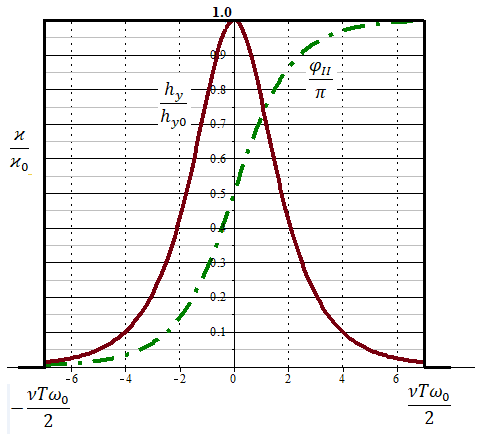}
\caption{Excitation of memory cell by time-aligned pulses of spin-polarized current (green) and magnetic field (red) for $\nu = 0.2$ and $M_{\textrm{s}} = 1.7\cdot 10^3$ G.}
\label{fig:mesh2}
\end{figure}

The word-line magnetic field pulse is taken to be time aligned with the pulse of the spin-polarized current flowing through the cell, as shown schematically in Fig. \ref{fig:mesh2}.

Similarly to the previous case, we assume that at $\tau<-\frac{T\omega_{0}}{2}$ the system was at its equilibrium, $\phi_{1}=0$. In turn, in the time interval $\tau \in \left[-\frac{T\omega_{0}}{2},\frac{T\omega_{0}}{2}\right]$, the magnetic moment of the free layer is affected by the field-current excitation such that angle $\phi_{\textrm{II}}$ can be found from
\begin{equation}
\begin{aligned}
\ddot{\phi}_{\textrm{II}}+(N_{y}-N_{x}) \sin(\phi_{\textrm{II}}) \cos(\phi_{\textrm{II}})-\kappa_{0} \cos(\phi_{\textrm{II}})\dot{\phi}_{\textrm{II}} \\ =h_{y} \cos(\phi_{\textrm{II}}).
\end{aligned}
\end{equation}

The limiting values at the moment $\tau_{1}$ of switching-on the excitation are equal to \(\phi_{\textrm{II}}=0\), \(\dot{\phi}_{\textrm{II}}=0\), but the presence of the right-hand side in (10) excludes the trivial solution, so \(\phi_{\textrm{II}}\neq0\). Thus, $\phi_{\textrm{II}}$ will be changing in time, and after switching off the current and field at \(\tau_{2}\), $\phi$ will continue to change described by
\begin{equation}
\ddot{\phi}_{\textrm{III}}+(N_{y}-N_{x}) \sin(\phi_{\textrm{III}}) \cos(\phi_{\textrm{III}})=0.
\end{equation}

Variables \(\phi_{\textrm{II}}\) and \(\phi_{\textrm{III}}\) at time \(\tau_{2}=\frac{T\omega_{0}}{2}\) are linked by the following boundary conditions:
\begin{equation}
\phi_{\textrm{II}}=\phi_{\textrm{III}}\mid_{\tau=\tau_{2}}; \dot{\phi}_{\textrm{III}}-\dot{\phi}_{\textrm{II}}=-\kappa_{0} \sin(\phi_{\textrm{III}})\mid_{\tau=\tau_{2}}.
\end{equation}

We will show that with a proper choice of the field excitation configuration, \(h_{y}(\tau)\) and pulse duration $T$, it is possible to achieve a regime of high-speed and essentially inertia-free switching of the free layer's magnetization from state \(\phi_{\textrm{I}}=0\) to state \(\phi_{\textrm{III}}\approx\pi\).

To do this, we assume that the configuration of the current pulse \(I_{x}\) and the corresponding time dependence of the magnetic field \(h_{y}\) (Fig. 2) are determined by a function, which can be approximated by
\begin{equation}
h_{y}=\frac{h_{y0}}{ch(\nu\tau)},
\end{equation}
where \(\nu\) is the coefficient determining the degree of temporal localisation of the magnetic field pulse, \(h_{y0}\) is the magnetic field amplitude.

The only requirement imposed on this parameter ($\nu$) is that \(\frac{\nu T\omega_{\textrm{0}}}{2}\gg1\). That is, at the boundary of the time interval $\left[-\frac{T\omega_{\textrm{0}}}{2},\frac{T\omega_{\textrm{0}}}{2}\right]$, the induced magnetic field is negligible. Therefore, the solution of equation (10), which describes the dynamics of the magnetization of the free layer, will be sought in the form
\begin{equation}
\phi_{\textrm{II}}(\tau)=2 \arctan(\exp{\nu\tau}).
\end{equation}

After substituting (14) into (10), the following identity is obtained:
\begin{equation}
(-\nu^{2}+\kappa_{\textrm{0}}\nu-(N_{y}-N_{x})+h_{y0})\frac{sh\nu\tau}{ch^{2}\nu\tau}=0.
\end{equation}
Relationship (15) defines the characteristic equation
\begin{equation}
\nu^{2}-\kappa_{\textrm{0}}\nu-h_{y0}+(N_{y}-N_{x})=0,
\end{equation}
from which coefficients \(\nu\) are found, allowing non-trivial solution (14) of equation (10).

It follows from (16) that \(\nu\) can have two values:
\begin{equation}
\nu_{\pm}=\frac{\kappa_{\textrm{0}}}{2}\pm\sqrt{(\frac{\kappa_{\textrm{0}}}{2})^{2}+h_{y0}-(N_{y}-N_{x})}.
\end{equation}

This result shows that switching can proceed via two regimes: fast, when the time parameter \(\nu=\nu_{\textrm{+}}\) and slow, when \(\nu=\nu_{\textrm{-}}\). In the special case where \((\frac{\kappa_{\textrm{0}}}{2})^{2}+h_{y0}-(N_{y}-N_{x})=0\), both processes become identical.

Analysing (17) makes it obvious that, for the magnetization reversal process to occur, the roots of \(\nu_{\pm}\) must have real values. This means that the contributions from the field and the spin-polarized current must exceed a certain threshold value determined by the shape anisotropy of the free layer:
\begin{equation}
    (\frac{\kappa_{\textrm{0}}}{2})^{2}+h_{y0} \geq N_{y}-N_{x}.
\end{equation}

At the same time, the magnitude of the magnetic field alone should not be sufficient to switch the free layer, since otherwise the induced magnetic field will lead to switching of all cells on the same word line. Therefore, the following conditions must be fulfilled to guarantee that only one selected cell is switched:
\begin{equation}
    h_{y0}<N_{y}-N_{x}<h_{y0}+(\frac{\kappa_{\textrm{0}}}{2})^{2}.
\end{equation}
The main role of the magnetic field \(h_{y}\) in this process is to bring the free layer's magnetization out of the collinear state within the memory stack.

As mentioned earlier, the process can proceed via fast or slow mode. The highest speed is achieved when \(h_{y0}\rightarrow{N_{y}-N_{x}}\) and \(\nu_{\textrm{+}}\rightarrow{\kappa_{\textrm{0}}}\). Therefore, this value corresponds to the upper limit of the magnetisation switching rate, and its magnitude is determined only by the amplitude of the spin-polarized current.

In turn, value \(\tau_{\textrm{0}}=\frac{1}{\nu}\approx\frac{1}{\kappa_{\textrm{0}}}\) can be considered as the characteristic duration of the system's magnetization reversal process. Therefore, a ``rectangular pulse'' of the spin-polarized current means in practice that its rise time should be much shorter than \(\tau_{\textrm{0}}\).

In order for expression (14) to be considered a solution to (10), (11), describing inertia-free switching of the magnetization of the free layer, it must satisfy the boundary conditions with high accuracy:
\begin{equation*}
    \phi_{\textrm{II}}=0, \dot{\phi}_{\textrm{II}}=0, \; \textrm{when} \; \tau=-\frac{T\omega_{\textrm{0}}}{2},
\end{equation*}
\begin{equation}
    \phi_{\textrm{II}}=\pi, \dot{\phi}_{\textrm{II}}=0, \; \textrm{when} \; \tau=\frac{T\omega_{\textrm{0}}}{2}.
\end{equation}
Here it is taken that \(\phi_{\textrm{I}}=0, \phi_{\textrm{III}}=\pi\).

Substituting (10) into the expressions for the boundary conditions (20) and taking into account the above discussed condition, (\(\frac{\nu T\omega_{\textrm{0}}}{2} \gg 1\)), we obtain
\begin{equation}
    \phi_{\textrm{II}}(\tau_{1})=2 \arctan(\exp{-\frac{\nu T \omega_{\textrm{0}}}{2}})\rightarrow{0},
\end{equation}
\begin{equation*}
    \phi_{\textrm{II}}(\tau_{\textrm{2}})=2 \arctan(\exp{\frac{\nu T \omega_0}{2}})\rightarrow{\pi},
\end{equation*}
\begin{equation*}
    \dot{\phi}_{\textrm{II}} (\tau_{\textrm{1}})=\dot{\phi}_{\textrm{II}} (\tau_{\textrm{2}})=\nu / \cosh(\frac{\nu T\omega_{\textrm{0}}}{2})\rightarrow{0}.
\end{equation*}

Thus, function (10), when the above conditions are met, describes with great accuracy the process of high-speed and almost inertia-free switching of the magnetization of the free layer from state \(\phi_{\textrm{I}}=0\) to state \(\phi_{\textrm{III}}=\pi\). In this regime, the post-switching magnetization oscillations are negligible.

It should be noted that the write process can be done in parallel on cells along the same word line in order to speed up the overall data storage speed. This would be in addition to the speed up of sequential writing of the same memory bit permitted by the relaxation-free (negligible post-switching oscillations) regime proposed herein.

We now estimate the duration of the spin-polarized current pulse, taking into account the parametric limitations discussed above. With $\nu T\omega_{\textrm{0}}/2=10 \gg 1$ and $\nu \approx \kappa_{\textrm{0}} = 0.2 \ll 1$, we obtain $T=\frac{2\cdot10}{0.2 \cdot \gamma \cdot 4\pi M_{\textrm{s}}} \approx 250$ ps, for $M_{\textrm{s}} = 1.7 \cdot 10^3$ G. The switching process considered in this work thus significantly speeds up the memory device in terms of single and multiple-sequential write operations. 

In conclusion, we present a method for ultra-fast inertia-free switching of a magnetic memory cell based on a double magnetic tunnel junction. The optimal time sequences and amplitudes of the field and spin-current pulses are determined for operation in the 100 ps regime, which should be attractive for high-performance MRAM.

\begin{acknowledgments}
Support from the National Academy of Sciences of Ukraine, the Swedish Research Council (VR 2018-03526), the Olle Engkvist Foundation (project 2020-207-0460), the Wenner-Gren Foundation (grant GFU2022-0011), and the Swedish Strategic Research Council (SSF UKR22-0050) are gratefully acknowledged.
\end{acknowledgments}

\bibliography{Refs}

\end{document}